\documentclass[10pts,aps,prx,twocolumn,superscriptaddress]{revtex4-2}
\usepackage{bbold}
\usepackage{amsmath}
\usepackage{amssymb}
\usepackage{xcolor}
\colorlet{rn}{red}
\usepackage{mathrsfs}
\usepackage{wasysym}
\usepackage{graphicx}
\usepackage{dcolumn}
\usepackage{bm}
\usepackage{lipsum}
\usepackage{braket}
\usepackage{epsfig}
\usepackage{graphicx}
\usepackage{orcidlink}
\usepackage{siunitx}
\definecolor{burgundy}{RGB}{97,0,35}
\definecolor{marine}{RGB}{4,46,96}
\definecolor{greendark}{RGB}{3,53,0}
\usepackage{hyperref}
\usepackage{bbm} 
\hypersetup{%
    colorlinks=true,
    urlcolor=burgundy,
    linkcolor=burgundy,
    citecolor=burgundy
    }

\usepackage[normalem]{ulem}

\newcommand{\Sr}[1]{$^{#1}$Sr} 
\DeclareSIUnit\gauss{G}
\newcommand{\StateThreePOne}{${^3\mathrm{P}_{1}}$}

\newcommand{\StateOneSZeroLong}{${\mathrm{5s^2} ^1\mathrm{S}_0}$}

\newcommand{\RedMotTransition}{${^1\mathrm{S}_0} \rightarrow {^3\mathrm{P}_1}$}
\newcommand{\ThreePTwoTransition}{${^1\mathrm{S}_0} \rightarrow {^3\mathrm{P}_2}$}
\newcommand{\RedBlastTransition}{$\ket{{^1\mathrm{S}_0},\, m_F = -9/2} \rightarrow \ket{{^3\mathrm{P}_1},\, F=11/2,\, m_F = -11/2}$}

\begin{document}

\DeclareGraphicsExtensions{.eps,.EPS,.pdf}

\title{Coherent control over the high-dimensional space \\of the nuclear spin of alkaline-earth atoms
}

\author{H. Ahmed\,\orcidlink{0009-0002-5061-1363}}
\affiliation{Laboratoire de Physique des Lasers, CNRS, UMR 7538, Universit\'e Sorbonne Paris Nord,  F-93430 Villetaneuse, France}

\author{A. Litvinov\,\orcidlink{0009-0001-5332-1188}}
\affiliation{Laboratoire de Physique des Lasers, CNRS, UMR 7538, Universit\'e Sorbonne Paris Nord,  F-93430 Villetaneuse, France}
\affiliation{Institut für Quantenoptik und Quanteninformation, Österreichische Akademie der Wissenschaften, 6020 Innsbruck, Austria}

\author{P. Guesdon\,\orcidlink{0009-0004-5809-6276}}
\affiliation{Laboratoire de Physique des Lasers, CNRS, UMR 7538, Universit\'e Sorbonne Paris Nord,  F-93430 Villetaneuse, France}

\author{E. Mar\'echal\,\orcidlink{0009-0000-3932-3331}}
\affiliation{Laboratoire de Physique des Lasers, CNRS, UMR 7538, Universit\'e Sorbonne Paris Nord,  F-93430 Villetaneuse, France}
\affiliation{Laboratoire de Physique et d’Étude des Matériaux, ESPCI Paris, Université PSL, CNRS, Sorbonne Université, F-75005 Paris, France}

\author{J. H. Huckans\,\orcidlink{0000-0001-6155-7476}}
\affiliation{ Department of Physical and Environmental Sciences,
Commonwealth University of Pennsylvania, Bloomsburg, PA 17815, USA}

\author{B. Pasquiou\,\orcidlink{0000-0001-5374-2129}}
\affiliation{Laboratoire de Physique des Lasers, CNRS, UMR 7538, Universit\'e Sorbonne Paris Nord,  F-93430 Villetaneuse, France}

\author{B. Laburthe-Tolra\,\orcidlink{0000-0002-5267-7334}}
\affiliation{Laboratoire de Physique des Lasers, CNRS, UMR 7538, Universit\'e Sorbonne Paris Nord,  F-93430 Villetaneuse, France}

\author{M. Robert-de-Saint-Vincent\,\orcidlink{0000-0002-8991-1366}}
\email{martin.rdsv@univ-paris13.fr}
\affiliation{Laboratoire de Physique des Lasers, CNRS, UMR 7538, Universit\'e Sorbonne Paris Nord,  F-93430 Villetaneuse, France}

\begin{abstract} 
We demonstrate coherent manipulation of the nuclear degrees of freedom of ultracold ground-state strontium 87 atoms, thus providing a toolkit for fully exploiting the corresponding large Hilbert space as a quantum resource and for quantum simulation experiments with  $\mathfrak{su}$(N)-symmetric matter. By controlling the resonance conditions of Raman transitions with a tensor light shift, we can perform rotations within a restricted Hilbert space of two isolated spin states among the 2F+1 = 10 possible states. These manipulations correspond to engineering unitary operations deriving from generators of the $\mathfrak{su}$(N) algebra beyond what can be done by simple spin precession. We present Ramsey interferometers involving an isolated pair of Zeeman states with no measurable decoherence after 3 seconds. We also demonstrate that one can harvest the large spin degrees of freedom as a qudit resource by implementing two interferometer schemes over four states.
The first scheme senses in parallel multiple external fields acting on the atoms, and the second scheme simultaneously measures multiple observables of a collective atomic state - including non-commuting ones. Engineering unitary transformations of the large spin driven by other generators than the usual spin-F representation of the $\mathfrak{su}$(2) group offers new possibilities from the point of view of quantum metrology and quantum many-body physics, notably for the quantum simulation of large-spin $\mathfrak{su}$(N)-symmetric quantum magnetism with fermionic alkaline-earth atoms. 
\end{abstract}

\date{\today}
\maketitle

\section{Introduction}
\label{Sec:Introduction}

Large-spin particles open new possibilities for quantum technologies. In the context of quantum many-body physics and quantum simulators, the enlarged spin degrees of freedom correspond to a significantly larger Hilbert space, which leads to qualitatively new physics \cite{Klempt2010pao, Krauser2012cmf, Taie2012asm, depaz2013nqm, Pagano2014aod, Patscheider2020cde, Alaoui2022mcf}. Large spins can also enhance quantum information processing \cite{Kiktenko2015sqr} and quantum metrology \cite{Fernholz2008sso, Hamley2012sns, Kajtoch2016ssi, Chalopin2018qes}. Conveniently, they emerge from several physical platforms, such as superconducting devices \cite{Peterer2015cad, Svetitsky2014htq}, molecular magnets \cite{Godfrin2018gri}, ions \cite{Randall2015epa}, multi-mode photons \cite{Lanyon2008sql}, and atoms \cite{StamperKurnsbg2013, Chomaz2022dpa}. 

Among large-spin atoms, fermionic alkaline-earth and alkaline-earth-like atoms (AEA) are particularly interesting for quantum metrology because they display a weak sensitivity to magnetic fields and possess narrow optical transitions that make them prominent for optical clocks with cold atoms \cite{Ludlow2015oac}. These atomic species are also remarkable for quantum many-body physics, as the purely nuclear nature of their spin in the ground state generates an $\mathfrak{su}$(N) symmetry that introduces frustration and should thus strongly impact quantum magnetism \cite{Gorshkov2010tos}. This possibility has attracted considerable attention from both theoretical \cite{Hermele2009mio, Corboz2012ssi, Cazalilla2014ufg} and experimental \cite{Taie2012asm, Hofrichter2016dpo, Ozawa2018asc, Tusi2022FSL} standpoints. To exploit the full potential of large-spin fermionic alkaline-earth atoms and to characterize the $\mathfrak{su}$(N) phases at low temperatures, it is crucial to develop tools that coherently control the spin states of atoms beyond simple spin precessions (see \cite{Leroux2018naa, Bataille2020, Barnes2022}).

 \begin{figure*}[t!]
\centering
  \includegraphics[width=1.5 \columnwidth]{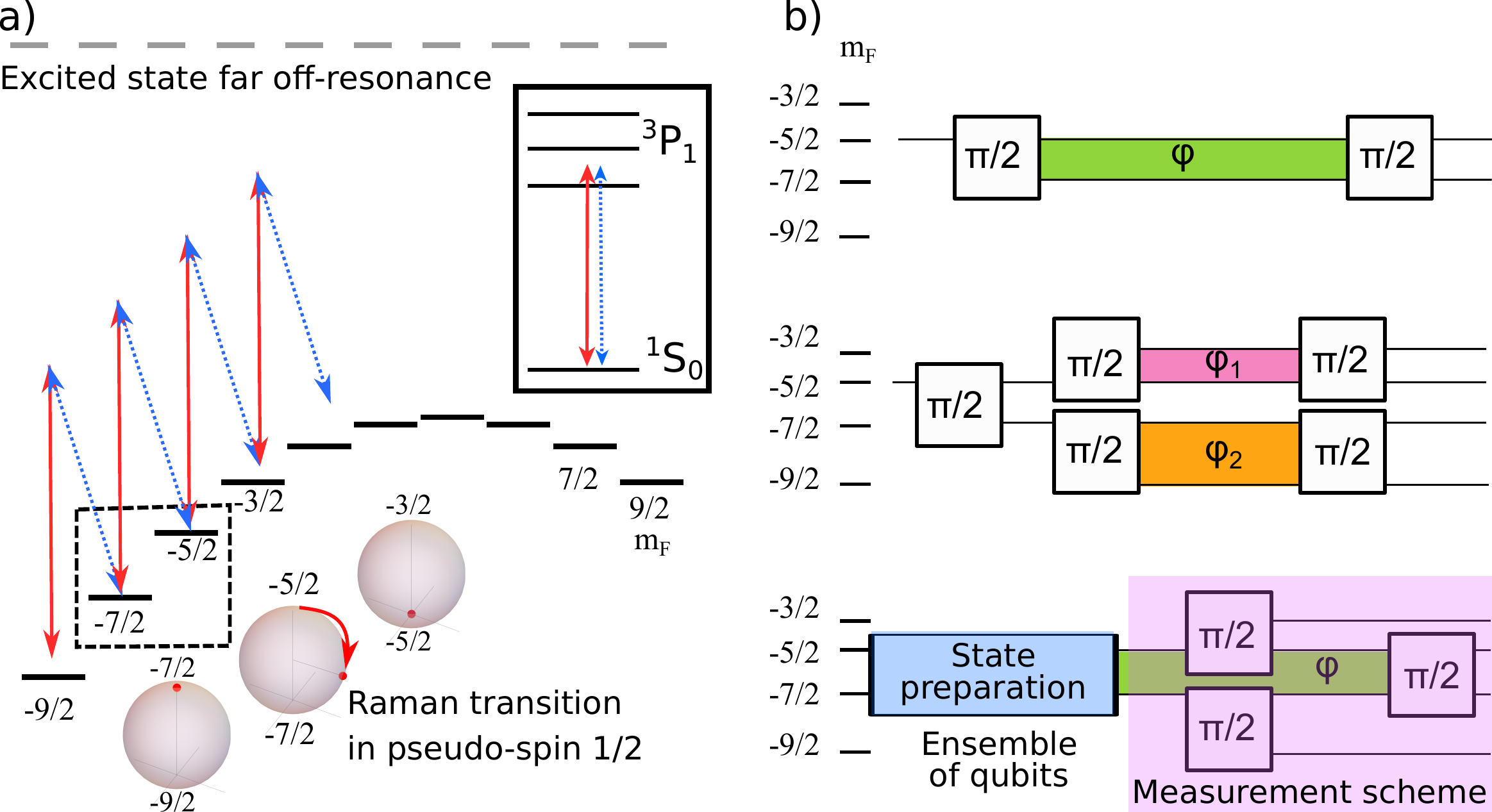}   
  \caption{\textbf{Principle of the experiment.} (a) Atomic levels and laser coupling scheme. A $\pi$-polarized light beam tuned within the hyperfine structure of the \RedMotTransition~transition (red arrows) creates an energy shift in the electronic ground state, with significant tensor component $U_{TLS}(m_F) = q \, m_F^2$. A second laser (here $\sigma^-$ polarized) enables Raman transitions. Thanks to the tensor light shift, the Raman resonance conditions differ from each other, making it possible to drive Rabi oscillations between two isolated spin states (here $\ket{-5/2}$ and $\ket{-7/2}$), all other processes being off-resonant. The Raman operation can be understood as a rotation in a sub-Bloch sphere. A potential linear energy shift $U_{linear}(m_F) = b \, m_F$ (in the illustration, $b \simeq - 3 q$) does not directly affect the capacity to isolate a pair of states. (b) The three experimental schemes in this paper. Top: We operate a Ramsey interferometer between states $\ket{-5/2}$ and $\ket{-7/2}$, decoupled from the other spin states. Middle: We operate two interferometers in parallel, demonstrating multi-parameter simultaneous sensing of the fields acting on the atoms. Bottom: we characterize the collective state of atoms in a superposition of two Zeeman levels ($\ket{-5/2}$ and $\ket{-7/2}$), identified as qubit states. The measurement sequence starts by applying two pulses that map the populations of the qubit states into ancillary states. Combining this with a final rotation in the qubit sub-Bloch-sphere, one can simultaneously measure two orthogonal components of the pseudo-spin of the ensemble of qubits.}
 \label{fig:fig1}
 \end{figure*} 

In this paper, we use a tensor light shift \cite{Deutsch2010qca} to control the resonance conditions of Raman transitions within the manifold of $2F+1=10$ Zeeman states of the spin-$F=9/2$ ground state of {$^{87}$Sr}. The quadratic energy shift $\propto m_F^2$ enables us to perform coherent manipulations between isolated pairs of $m_F$ states with a fidelity better than {99\,\%}. These operations demonstrate control over unitary transformations deriving from $\mathfrak{su}$($N = 2F+1$) generators, beyond the usual precession operations deriving from the spin-F representation of the $\mathfrak{su}$(2) group. Taking advantage of this new possibility and the atoms' insensitivity to magnetic fields, we operate a Ramsey interferometer between two isolated hyperfine states and demonstrate Ramsey fringes with no discernible loss of contrast over 3\,s. We further exploit the large spin of the atoms with two new types of interferometers. In the first implementation, we operate two Ramsey interferometers simultaneously and in parallel, using two independent pairs of spin states within the hyperfine structure. This configuration allows simultaneous measurement of two independent parameters (here corresponding to the effective quadratic and linear Zeeman shifts). In the second implementation, we mirror two physical spin states that make a qubit into two ancillary spin states. Selective unitary transformation (``rotation") of the two qubit states and the measurement of all four states then permit the simultaneous measurement of two non-commuting observables of an ensemble of qubits.

\section{Implementation of $\mathfrak{su}$(N) generators.}
\label{Sec:ExperimentPrinciple}

One key idea of this paper is to implement coherent rotations between two and only two arbitrary spin states chosen among the $N$ spin states of an $N-$component Fermi mixture. Rabi oscillation between such a selected pair of states corresponds to rotations driven by infinitesimal generators of the $\mathfrak{su}$(N) spin algebra, described by an operator $\sigma^x_{m_F,m_F'}$. The restriction of $\sigma^x_{m_F,m_F'}$ to the Hilbert space $\{\ket{m_F} , \ket{m_F'} \}$ is the Pauli matrix $\sigma^x$, while all its other elements are zero. The evolution operator after a sequence reads  $\sim \exp{(- i \theta \sigma^x_{m_F,m_F'}/2 )}$, where $\theta = \int \Omega_R(t) dt$ is called pulse area and $\Omega_R$ is the Rabi frequency. The evolution operator describes a rotation in the restricted Hilbert space associated with the two selected spin states, and is the identity in the orthogonal space consisting of all other levels. Thus, it is distinct from the more usual manipulation of a spin $\vec F$ by precession around a magnetic field, $\exp(-i \vec \alpha . \vec F)$, which involves a combination of the three generators $\vec F = (F_x, F_y, F_z)$ of the spin-F representation of the $\mathfrak{su}$(2) algebra. This new tool enables complete control over the spin degrees of freedom, which is necessary to characterize phases of $\mathfrak{su}$(N)-symmetric quantum matter at low energy (see e.g. \cite{Chen2005esp}) and for computing applications leveraging the vast Hilbert space provided by large spins (see e.g. \cite{Kiktenko2015sqr}).

It is challenging to intuitively represent the state of large-spin atoms and the $\mathfrak{su}$(N) generators of rotations we implement since a spin $F>1/2$ possesses a rich inner structure that the usual Bloch sphere fails to represent fully. This structure can be captured by, for example, the Majorana stellar representation \cite{Majorana1932aoi} (see also \cite{Barnett2006cnp, Bruno2012qgp, Evrard2019ems}). However, the motion of the Majorana stars driven by coherent manipulations is generally quite complicated and does not always allow a clear physical interpretation. Qualitatively, $4F$ angles ($2F$ points on a sphere) are sufficient to describe the state. A simple representation, that can be well-suited to visualize some of our interferometric experiments, is to consider each two-level sub-space of adjacent Zeeman levels ($m_F, m_F+1$) in a Bloch sphere representation and consider the set of $2F = N-1$ Bloch spheres as a representation of the state. For each Bloch sphere, the latitude represents the relative populations while the longitude represents the relative phase of the spin wavefunction in the two relevant states. 

Experimentally, we perform rotations on a sub-Bloch sphere between selected pairs of nuclear-spin states, either $m_F \longleftrightarrow m_F' = m_F+1$ or $m_F \longleftrightarrow m_F' = m_F+2$ - see Fig.~\ref{fig:fig1}. To this end, we introduce two-photon Raman couplings. We rely on a quadratic dependence of the light shift as a function of $m_F$ (tensor light shift, TLS), to resonantly drive only one of the two-photon resonances, whereas the other couplings are off-resonant and have negligible impact on the atomic state evolution. 
Provided the Raman coupling is much smaller than the TLS, we can spectrally resolve any pair of $m_F$ states (with $\Delta m_F = 1,\,2$), beyond what was achieved by the approach in Ref.~\cite{Barnes2022}, which only allowed coherent oscillations between $m_F = F$ and $m_F = F-1$ (or $m_F = -F$ and $m_F = -F+1$).
Our work is thus analogous to the proposal in Ref.~\cite{Omanakuttan2021} but with Raman couplings instead of radio-frequency couplings between adjacent Zeeman sublevels. It is also very similar to the experiment on the solid-state platform in Ref.~\cite{Godfrin2018gri}, where a hyperfine interaction, instead of an external field, lifts the degeneracy.  

Pulsing the Raman coupling drives precession on one selected sub-Bloch sphere around the x axis (by phase convention). Furthermore, at any time, all of the sub-Bloch spheres experience $\sigma_z$ precession at different rates determined by the detuning of the Raman beat note (controlled by an RF local oscillator) to the corresponding two-level transition. In this article, we explore three independent interferometric schemes restricted to a given Hilbert subspace (see the right panels of Fig.~\ref{fig:fig1}). They illustrate the new possibilities for quantum sensing or simulation that emerge when using simple combinations of coherent rotations between pairs of spin states, complementing previously demonstrated schemes using large-spin atoms \cite{Fernholz2008sso, Chalopin2018qes, Evrard2019ems}).

The intercombination line \RedMotTransition~of the \Sr{87} isotope is particularly favorable to engineer large tensor shifts with only weak spontaneous emission, thanks to the high ratio of about $10^5$ between the hyperfine splitting in $^3P_1$ and the inverse lifetime of these hyperfine states \cite{Burba2024}. Using the much narrower \ThreePTwoTransition~transition might lead to even better performances. Our scheme should readily adapt to other alkaline-earth-like atoms and to other species with similarly rich electronic structures (Yb, Cd, Hg, Er, Dy, ...). Finally, compared to the recent results in Ref.~\cite{claude2024, Zheng2024}, where the spins of erbium and strontium atoms, respectively, are manipulated through coherent oscillations on a narrow optical transition, our scheme also benefits from \textit{not} relying on a clock-like optical transition. This, we achieve because the Raman transitions negligibly populate lossy excited states.

\section{Experimental setting}
\label{Sec:ExperimentSetting}

All the experiments described in the following sections start with a spin-polarized ultracold Fermi gas of \Sr{87}. To produce this gas, we first laser cool and hold within a dipole trap a $\SI{5}{\micro \kelvin}$-cold sample of atoms in the \StateOneSZeroLong~ground state, fully depolarized across the ten spin states of the $F = 9/2$ manifold. We then use optical pumping on the {\RedMotTransition, $F = 11/2$} intercombination line to empty all spin states but two (typically, the $m_F = -9/2$ and $m_F = -5/2$ states). We thus prepare a stable spin mixture since the $\mathfrak{su}$(N) symmetry prevents spin-changing collisions. Ref.~\cite{Bataille2020} gives more details on this optical pumping. We next use forced evaporative cooling to obtain an ultracold gas at a temperature of about $\SI{100}{\nano \kelvin}$. Finally, we apply a state-selective radiation-pressure pulse to remove atoms in the $m_F = -9/2$ state, with no noticeable heating of the remaining $m_F = -5/2$ sample. We achieve such state selection by a 4.5 ms light pulse tuned to the \RedBlastTransition~transition in a magnetic field of $ \SI{4.5}{\gauss}$. The polarized gas, of about $10^4$ atoms, is then kept in a uniform magnetic field $B \vec e_z$ with $B = \SI{5.2}{\gauss}$ that defines the quantization axis and produces a linear Zeeman splitting of $960 (5)$ Hz. With less than 1$\%$ of the atoms in each Zeeman state other than the one chosen, the residual spin entropy is below 0.5 (while a fully random spin state has entropy log(10) = 2.3). 

After this preparation, we coherently manipulate the atoms' nuclear spin using an external-cavity laser diode whose frequency is tuned within the hyperfine structure of the \StateThreePOne~state, $\SI{600}{\mega \hertz}$ to the red of the {$\ket{^1S_0, F=9/2} \rightarrow \ket{^3P_1, F=9/2}$ }~line. We produce two phase-coherent laser beams with independently controlled frequencies from a single laser split into two paths with independent single-pass acousto-optic modulators. The first beam's primary role is to engineer a quadratic energy shift $U_{TLS}(m_F) = q \, m_F^2$ of the Zeeman states labeled by their spin projection $m_F$. We will refer to this beam as the ``Tensor Light Shift (TLS)" beam. The second beam, labeled the ``Raman" beam, acts together with the first (in most of the paper) to drive Raman transitions between two Zeeman states when their frequency difference matches the energy difference between states. We drive the AOMs using two AD9852 direct digital synthesizers synchronized on a shared $\SI{20}{\mega \hertz}$ clock. The beam frequency difference thus has a long term relative instability directly set by the shared clock (RIGOL DG1022), at the {5\,ppm} level. They are recombined on the same path with orthogonal linear polarizations through a single-mode optical fiber before being sent onto the atoms with a beam waist of $\SI{320}{\micro \meter}$. This configuration also renders recoil effects negligible. 

We set the TLS beam polarization almost co-linear to the magnetic field, corresponding to $\pi$ transitions in this quantization axis. We ramp its power up to typically $\SI{5}{\milli \watt}$ in $\SI{3.5}{\milli \second}$, slowly enough that the atoms' spin state adiabatically follows the slight change of quantization axis. We stabilize the beam intensity to a sub-percent level at the millisecond timescale, using a photodiode positioned after the beam has crossed the atoms. The Raman beam polarization is linear and orthogonal to the magnetic field, allowing $\sigma^+$ and $\sigma^-$ transitions. Throughout this article, we mainly apply $\delta m_F = 1$ spin-changing Raman transitions from absorption and stimulated emission of one $\pi$ photon from the TLS beam and one $\sigma^-$ photon from the Raman beam. By setting the two-photon detuning resonant with such transitions, the $\sigma^+$ Raman beam photons only drive widely off-resonant processes thanks to the TLS beam's quadratic shift $U_{TLS}(m_F)$ and ultimately remain spectators. Moreover, the $\sigma^-$ transitions addressed by the Raman laser have higher Clebsch-Gordan coefficients and coupling strengths when manipulating $m_F <0$ states. In an alternative scheme, the Raman laser is bichromatic, and drives alone $\delta m_F = 2$ two-photon spin-changing transitions using its two circular polarization components. The two optical frequencies are generated on the same AOM by a bichromatic RF drive, produced by frequency mixing.

At the end of an experiment cycle, we adiabatically ramp down the TLS beam and measure the spin-state distribution by spin-selective momentum transfer \cite{Bataille2020}, simultaneously extracting and probing two spin states of choice ($m_F-1$ and $m_F+1$) for each measurement. Spin populations presented throughout the paper are relative to the total atom number. For this, we calibrate the detection efficiencies $\eta(m_F)$ for each spin state on clouds prepared in that single spin state. For  $m_F = -7/2$, $-5/2$, and $-3/2$, we find $\eta(m_F) = $ 0.65, 0.70, and 0.51, respectively\,\footnote{The detection efficiencies are probably limited by a lack of optical power and by using rather warm clouds with momentum spread comparable to the optical recoil.}. However, $\eta(-3/2)$ fluctuated throughout the experiments shown here, as evidenced by some population estimates larger than one. In such cases, we recalibrate $\eta(-3/2)$ by up to $6\,\%$ to bring extremal population estimates down to 1.

\section{Coherent manipulation of a nuclear-spin qubit}
\label{Sec:CoherentManipulation}

 \begin{figure*}[t!]
\centering
  \includegraphics[width= 2 \columnwidth]{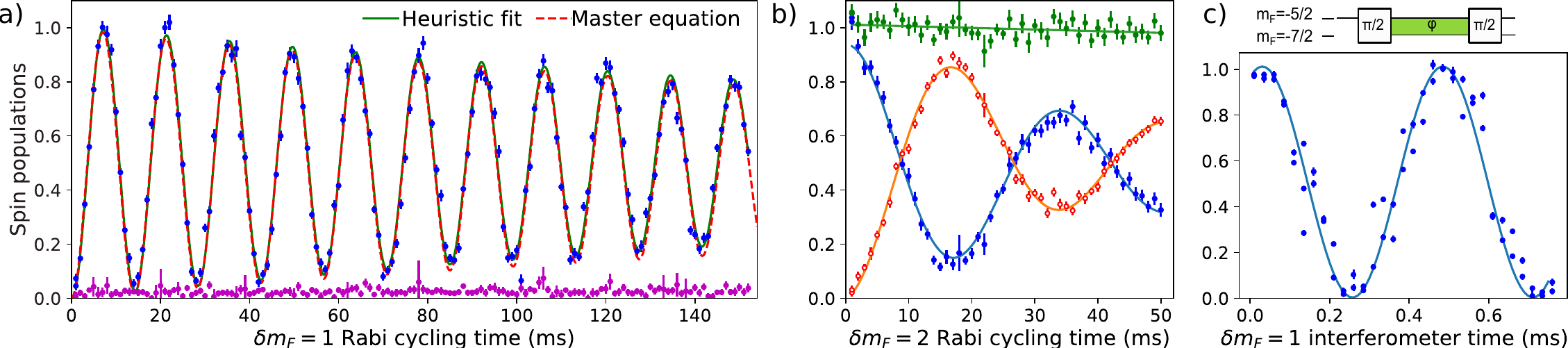}
   \caption{\textbf{Encoding a qubit on two nuclear spin states.} 
   a) Rabi oscillations between $m_F=-3/2$ (blue) and $m_F=-5/2$ (not shown), driven by a $\delta m_F = 1$ Raman coupling. Here $\hbar \Omega_R / 2q$  is small enough that the neighboring states $m_F=-7/2$ (magenta) and $m_F= -1/2$ (not shown) are unaffected. The green line is a heuristic fit to the data with a damped sine, yielding a $1/e$ damping time of 298\,(17) ms. The dashed red line is the result of a master equation simulation with two adjustable parameters: the Raman coupling strength and the coherences decay rate. It is almost indistinguishable from the damped sine. (b) Rabi oscillations between $m_F=-3/2$ (blue, filled circles) and $m_F=-7/2$ (red, empty circles) driven by a $\delta m_F = 2$ Raman coupling. Green dots show the total population in these two states. Both populations are measured simultaneously. Solid lines are fits with damped sinusoidal oscillations, yielding a $1/e$ damping time of {40(5)\,ms}.
   c) Ramsey interferogram realized between states $m_F=-7/2$ (blue dots) and $m_F=-5/2$, driven by a $\delta m_F = 1$ Raman coupling. The line is an optimal distance regression fit with a sinusoidal interferogram. In all of this figure, each data results from a single shot, and error bars reflect the population measurement statistical uncertainty. Furthermore, in plots (a) and (b), the detection efficiencies $\eta(-3/2)$ are slightly recalibrated to bring maximal $m_F = -3/2$ population estimates down to 1.
   }
 \label{fig:two-level-effects}
 \end{figure*}

\textit{Rabi oscillations between two isolated hyperfine states}. We now turn to the experimental demonstration of coherent manipulations between selected pairs of states within the \Sr{87} ground state manifold, which realizes evolutions driven by $\mathfrak{su}$(N) generators  $\sigma^x_{m_F, m_F'}$ and $\sigma^z_{m_F, m_F'}$. We first show Rabi oscillations between two Zeeman sublevels, then characterize the coherence of quantum superpositions by Ramsey interferometry.

We first demonstrate Rabi oscillations with $\delta m_F = 1$. To this end, we prepare the gas in the $m_F = -5/2$ state and turn the quadratic light shift on $U_{TLS}(m_F) = q m_F^2$. For $\delta m_F = 1$ Raman transitions, this splits the resonance conditions of the various Raman processes by multiples of $2 q /h $ (h is the Planck constant). Consequently, as illustrated in Fig.~\ref{fig:fig1}, tuning the Raman beam frequency to one of the resonance conditions (e.g., the process $-5/2 \rightarrow -3/2$) produces a state evolution that can be approximated as generated by a Pauli operator coupling two and only two states. To reach this regime, the separation between adjacent resonances $2 q /h$ (proportional to the TLS beam intensity $I_{TLS}$) must be larger than the two photon Raman coupling strength ($\Omega_R \propto \sqrt{I_{TLS} I_R}$, where $I_R$ is the Raman beam intensity), which implies $I_R \ll I_{TLS}$. In practice we use $I_R / I_{TLS} \le 10^{-4}$. 

In the presence of the TLS beam, we pulse the Raman beam at a fixed frequency and monitor the spin state evolution. In Fig.~\ref{fig:two-level-effects}a, we demonstrate the Rabi oscillation dynamics driven between the states $m_F = -5/2$ and $m_F = -3/2$. Here $q / h = -\SI{320}{Hz}$ and $\Omega_R / 2 \pi = \SI{71}{\hertz}$. We observe well-pronounced coherent oscillations with a short-time contrast consistent with 1, while the population in the neighboring, not targeted state $-7/2$ grows by less than $1\,\%$ over almost 11 Rabi oscillations. The amplitude of the oscillations exhibits a slow exponential decrease with a characteristic time $\tau_{decay} = \SI{298 \pm 17}{\milli \second} \simeq 21 \frac{2 \pi}{\Omega_R}$. As we do not observe significant population diffusion to other states, this damping rate limits the fidelity of a $\pi/2$ pulse to 0.994. The shot-to-shot fluctuations of the Rabi pulse area away from $\pi/2$ (calibrated from complementary experiments) are even less limiting to the fidelity ($> 0.998$). This demonstration of an evolution well approximated by $\exp{\left( - i \theta \sigma^x_{m_F, m_F+1} /2\right)}$ constitutes a key result of this article, enabling the various experiments described in the following sections.  

We also demonstrate Raman transitions with $\delta m_F = 2$ using a slightly different scheme: we detune the Raman beam 1 MHz below the TLS beam and modulate its frequency in two sideband components (suppressing the carrier frequency). In this manner, its two opposed $\sigma^\pm$ polarization components drive $\delta m_F = 2$ transitions, with a detuning controlled by the modulation frequency. We performed this experiment at a much lower $q/h = -95\,$Hz (because of the aging of a laser component). In Fig.~\ref{fig:two-level-effects}b, we observe the Rabi oscillation induced between the states $-7/2$ and $-3/2$. The two spin populations, monitored in parallel, exhibit complementary oscillations at frequency {$\Omega_R/ 2 \pi = 29\,(3)\,$Hz} with a damping time of {$40\,(5)$\,ms}. The total population in the two states decays by 3$\%$ within the damping time, so the damping mainly reflects a loss of coherence, not diffusion to other spin states. The limit imposed by this damping on the fidelity of a $\pi/2$ pulse can thus be estimated at 0.90, much lower than what we achieve on the $\delta m_F = 1$ Raman processes. The cause of this more significant damping compared to the $\delta m_F = 1$ coupling experiment is unknown - possibly the laser degrading, which also resulted in the lower value of $q$.
Still, this experiment demonstrates a complementary set of two-level rotations. They are not required to have complete control over the atomic state, but, in principle, an extended set of generators of evolution can be advantageous for some manipulations (assuming similar fidelities). Furthermore, combining $\delta m_F = 1$ and $\delta m_F = 2$ transitions enables conceptually new kinds of experiments, such as engineering dynamics in a spin space with periodic boundary conditions, with phenomena related to Berry phases and non-Abelian gauge fields \cite{Ruseckas2005, Leroux2018naa}. In the remainder of the paper, we will use exclusively $\delta m_F = 1$ transitions as the basic building block for interferometers. 

\textit{Coherence of the Rabi oscillations}. Let us characterize the coherence properties of the nuclear qubit - first in the presence, then in the absence, of the Raman driving field. In the Rabi oscillations presented in Fig.~\ref{fig:two-level-effects}a, each point is the outcome of an individual realization. The observed damping is seen without averaging different experimental realizations, which implies that it is associated with loss of coherence rather than fluctuations. This decoherence may result from a spatial inhomogeneity or from spontaneous emission. To establish which one dominates, we measure the heating rate of atoms exposed to TLS light in the dipole trap. We find it to be three times higher than the expected heating rate associated with spontaneous emission assuming the laser is monochromatic. We attribute this to a frequency noise pedestal in our laser spectrum due to amplified spontaneous emission (ASE) that causes \textit{resonant} photon scattering. We then model the Rabi dynamics by a master equation describing the evolution of an atom in the laser and magnetic fields. We calculate the separate spontaneous emission terms from and to each spin state, such that the model captures decoherence and spin-state diffusion. To heuristically capture the effect of ASE, we scale these rates of coherence decay and spin diffusion by a factor of three. Typical coherence decay rates are $\sim 1\,$s$^{-1}$, and population transfer rates $\sim 0.5\,$s$^{-1}$. However, the simulation with these parameters does not produce the evolution observed in Fig.~\ref{fig:two-level-effects}a, nor does artificially scaling the impact of spontaneous emission by a different factor. Instead, we note that there is TLS/Raman beam inhomogeneity, and empirically add a decay of coherences between spin states $(m_1, m_2)$ with rate $\Gamma_q \times |m_1^2-m_2^2|$. With $\Gamma_q = 1.2\,\rm{s}^{-1}$, we recover excellent agreement with the data, as shown in Fig.~\ref{fig:two-level-effects}a. We estimate independently, by simulations with a random distribution of $(q,\Omega_R)$, that the observed dynamics is consistent with an inhomogeneity of $q$ and $\Omega_R$ of about $1.2\,\%$ in standard deviation. Across a 5$\mu$m-radius cloud, this can be induced by a beam off-centered by $w/4$, where $w=320\,\mu$m is the waist.

\textit{Coherence of the hyperfine qubit Ramsey interferometer}. We then test the coherence of the qubit in the absence of Raman drive. For this, we implement an interferometer. Keeping the TLS beam on, we apply a set of two resonant $\pi/2$ pulses, separated by a temporal interval $T$ - thus realizing a standard Ramsey sequence restricted to the Hilbert space of a qubit, see the top sequence schematic of Fig.~\ref{fig:fig1}b. During this time interval $T$, we alter the local oscillator frequency that controls the phase difference between TLS and Raman beams. The Ramsey interferometer is sensitive to the phase $\phi$ defined by $\frac{d \phi}{dt} = \Delta E_{m_F,m_F'} (t)/\hbar - \delta_{RF} (t) $, where $\Delta E_{m_F,m_F'}$ is the energy difference between the two qubit states and $\delta_{RF} (t) = \omega_R-\omega_{TLS}$. The quantities $\omega_{TLS} / 2 \pi$ and $\omega_R / 2 \pi$ are the (potentially time-dependent) TLS and Raman beam laser frequencies.

Figure~\ref{fig:two-level-effects}c shows the fringe pattern as a function of $T$ for qubits encoded in the two nuclear-spin states $m_F = -5/2$ and $m_F = -7/2$. Here, $q / h = -\SI{190}{\hertz}$ and $\Omega_R / 2 \pi = \SI{93}{\hertz}$. The well-resolved oscillations come from the accumulated differential phase in the qubit states, resulting in a rotation (controlled by $T$) along the sub-Bloch sphere equator. We observe fully contrasted oscillations for the best data,  which demonstrates that the coherence of the qubit is well preserved. 

\begin{figure}[t!]
\centering
 \includegraphics[width=\columnwidth]{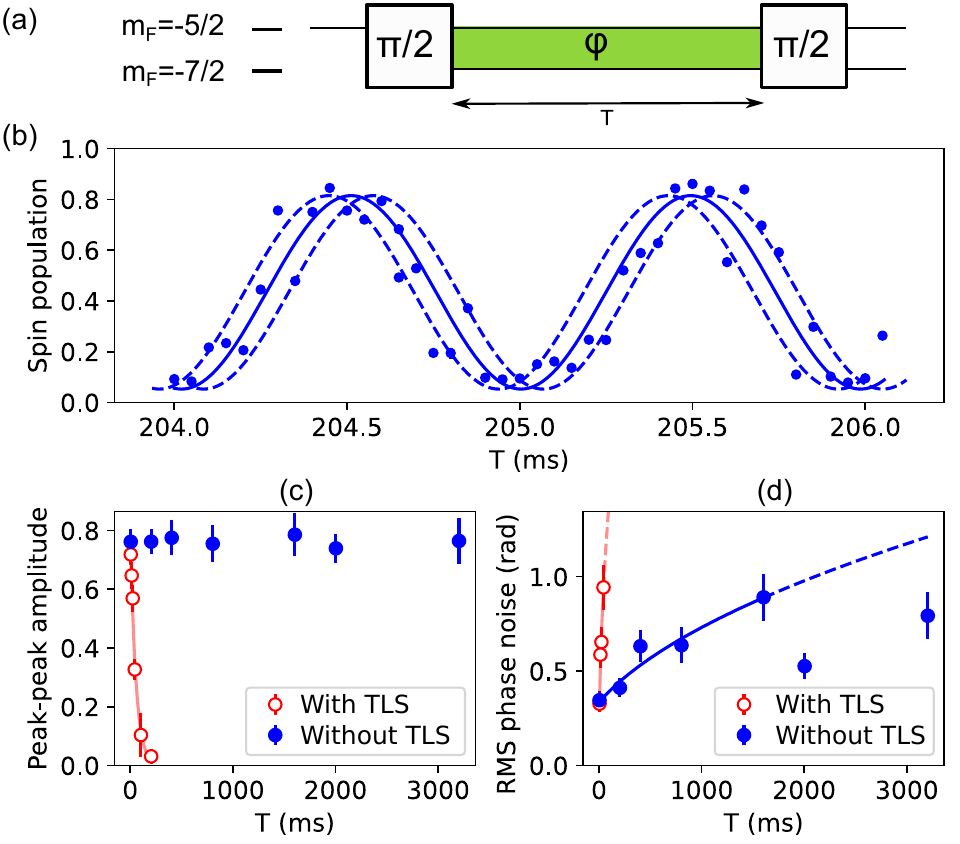} 
  \caption{\textbf{Qubit coherence.} 
  (a) Schematics of the interferometer. (b) Fringe pattern of a Ramsey interferometer operated between $m_F = -5/2$ (measured) and $m_F = -7/2$, with, between the two pulses, {4\,ms} of exposition to TLS light and {$T - 4\,$ms} of absence of any laser light. Error bars, combining population measurement uncertainty and expected impact of pulse area fluctuations, are comparable to the dots' size. The solid line is the underlying interferogram, barring phase fluctuations. The dashed lines interferograms are shifted by $\pm$ one standard deviation of the phase fluctuations. (c) Contrast as a function of $T$. The red line is a fit to an exponential decay for the data with TLS beam on (red, empty circles), with $1/e$ decay time of {$54(5)$\,ms}. When the TLS beam is adiabatically turned off during the dark time (blue, filled circles), there is no observable contrast decay for at least {3\,s}. The shortest measured $T$ ({5\,ms}) corresponds to ramping down and almost immediately back up the TLS. (d) Phase noise of the interferometer as a function of $T$. Continuous lines are fits assuming $Var(\phi) = \delta\phi_0^2 + D t$. We have $\sqrt{D} = 0.14(1)\,$rad/ms$^{1/2}$ with TLS, and $\sqrt{D} = 0.021(2)\,$rad/ms$^{1/2}$ without TLS.}
 \label{fig:qubit_coherence}
 \end{figure}

We now systematically study  the contrast and noise of Ramsey interferometers as a function of $T$ to characterize the sources of decoherence. We will highlight that the main detriment to the qubit's coherence is the TLS beam and that we can adiabatically remove it to recover coherence over seconds. We present this study in Fig.~\ref{fig:qubit_coherence}. It is realized with approximately the same parameters as the study of coherence in Rabi oscillations: $q/h =  -\SI{300}{\hertz}$, $\Omega_R/2 \pi = \SI{72}{\hertz}$. We focus on two effects: decoherence, which reduces the interferometric contrast, and phase fluctuations. In principle, the sources of fluctuations in the data are (i) atomic quantum projection noise; (ii) measurement uncertainties associated with fitting the spin-separated optical density pictures; (iii) reproducibility of the $\pi/2$ pulses; iv) fluctuations of the phase integrated by the interferometers. For our data, the quantum projection noise (i) is smaller than the measurement noise (ii) by typically a factor of 3, so we neglect it. Fluctuations of the area of ``$\pi/2$" pulses (iii) are estimated from repeating a sequence without closing the interferometer. We find a standard deviation of the pulse area of 0.063(5) rad (which sets a fidelity upper bound for a $\pi/2$ pulse at 0.999). However, their effect in a \textit{closed} Ramsey sequence remains smaller than the measurement uncertainty. On the interferogram shown in Fig.~\ref{fig:qubit_coherence}b, we notice that the residuals to a sinusoidal model are much larger than the fluctuations expected from combining the effect of (i), (ii), and (iii). We attribute this to uncontrolled fluctuations of the interferometer phase. To estimate their RMS amplitude, we numerically generate trial noisy models of the experiment, including all sources of fluctuations, and compare them with the data\,\footnote{We look for a trial model for which a least-square sinusoidal fit estimates the same amplitude and with the same residuals as the data. Note that the peak-to-peak amplitude of the sinusoidal fit is generally smaller than the contrast of the interferogram: an interferogram with complete loss of control over the phase will be fitted with zero amplitude and large residuals.}. In this optimization, the interferometric contrast and the variance of phase fluctuations are the adjustable parameters.

Fig.\,\ref{fig:qubit_coherence}c and d show the outcome of this analysis. As shown in Fig.\,\ref{fig:qubit_coherence}c, the contrast (defined as peak-to-peak signal amplitude) decays as a function of $T$ with a $1/e$ decay constant of {$54\,(5)$\,ms}. Like the damping of Rabi oscillation, this decay most likely results from the TLS inhomogeneity rather than photon scattering. The observed decay is significantly faster than that of the Rabi oscillations seen in Fig.\,\ref{fig:two-level-effects}a. This illustrates the sensitivity of Ramsey interferometers, which are directly sensitive to detunings $\Delta$, in contrast to generalized Rabi frequencies, generically $\sqrt{\Omega_R^2+\Delta^2}$.

A second observation is that a random phase is integrated for each realization of the interferometer. At the smaller dark time {$T = 5\,$ms}, its standard deviation is about {0.35\,rad} and gradually increases with $T$, as shown in Fig.\,\ref{fig:qubit_coherence}d. It is about {1.4\,rad} after {100\,ms}. We attribute this to polarization fluctuations in the TLS beam, as we will demonstrate in Sec.~\ref{SubSec:ParallelRamsey}. One could reduce these fluctuations by polarization filtering, although our setup, where we co-propagate the TLS and Raman beams in the same optical fiber, is inadequate for this. 

Thus, simply removing the tensor light shift for most of the duration $T$ of the interferometer should significantly mitigate decoherence and phase noise. After the initial Rabi pulse, we adiabatically turn off the TLS in {2\,ms}, and bring it back on in the last {2\,ms} before the second $\pi/2$ pulse. The atom's environment during the interferometer is now primarily the magnetic field and a dipole trap with negligible spin-dependent light shift. The observed decoherence is considerably suppressed, with no visible contrast decay after 3 seconds. The phase noise also grows with $T$ at a much slower rate. Phase fluctuations of $\simeq 0.3$ rad are acquired during the {4\,ms} with TLS, but the phase-noise contribution of the remaining time without light only reaches 1 radian after about 2 seconds. Although the RMS phase fluctuations are not exactly linear with $T$, they can qualitatively be interpreted as an instability of the Raman detuning of {$\sim 0.08\,$Hz} when integrated (averaged) over two seconds. The frequency instability of the DDS-controlled $\delta_{RF} = \omega_R - \omega_{TLS}$ is, in principle, one order of magnitude smaller. Thus, the phase noise probably reflects magnetic field fluctuations of about {0.4\,mG} at the second timescale, consistent with the specifications of the current supplies for the coils. 

In this section, we have demonstrated coherent control of a qubit defined over two selected Zeeman sublevels using $\sigma^x_{m_F, m_F+1}$, $\sigma^x_{m_F, m_F+2}$, and $\sigma^z_{m_F, m_F'}$ rotations. Our demonstrations have been restricted to the manifold $m_F \in \{ -9/2, -7/2, -5/2, -3/2 \} $. Note that, when driving atoms in other states, we observed uncontrolled population transfers. We attribute this to the value of our magnetic field $b \simeq -3q$, which causes quasi-degeneracy and sign reversal of the Raman energy differences, see Fig.~\ref{fig:fig1}a. One could lift this issue and acquire complete control over the entire spin manifold, provided $b > |q| (2F-1)$. Within the manifold of states used in this paper, we have shown the high sensitivity of the qubits' coherence and phases to the TLS beam 
fluctuations. Minimizing the exposure time to the TLS beam, we have demonstrated long-lived coherent superpositions over several seconds, without any magnetic shielding. Such coherence time is a feature of the minute Land\'e factor of the nuclear spins, basically immune to stray magnetic field gradients ($g_I \mu_0 \nabla B < 10^{-4}\,\rm{Hz}/\mu$m), and of the extremely small vector and tensor polarisabilities in the closed-shell ground state, at the optical trapping wavelength (with spin-dependent shifts $\simeq 10^{-4}$\,Hz). It can be a great asset, e.g., for high-precision sensing in long-interrogation-time interferometers or quantum information storage.  

\section{Beyond pseudo-spin 1/2 schemes: qudit interferometry}
\label{Sec:BeyondPseduoSpin}

The possibility to coherently couple two spin states among the $N=10$ available ones in ground-state $^{87}$Sr atoms enables coherent navigation over the whole ground-state manifold - for example, by simply concatenating a series of resonant pulses between isolated pairs of Zeeman states. Consequently, our experimental toolkit creates new possibilities that use more than two spin states in order to store or retrieve information. In this section, we experimentally investigate two of those new possibilities that use the large nuclear spin as a genuine qudit.

First, we will use well-defined linear-superposition states involving four spin states, in order to simultaneously measure more than one external field at each experimental realization. Such multi-parameter estimates are useful in the context of metrology \cite{Baamara2023qem}. Our work thus illustrates the potential of large-spin particles as original quantum sensors.

Second, we present a scheme that simultaneously measures collective observables that do not commute, in a single experimental realization. Building on Ref.~\cite{Kunkel2019}, our scheme can characterize a many-body state where the coherences are stored in a subset of the fully available ground-state hyperfine manifold. The main idea is to extend the Hilbert space of the atoms for measurement purposes only; the measurement sequence involves mapping the state of the atoms into initially empty ``ancillary" states and applying controlled rotations, which provides additional observables, and facilitates tomographic characterization of a quantum many-body state.

\subsection{Parallel Ramsey interferometers for simultaneous multiple-field sensing.}
\label{SubSec:ParallelRamsey}

Performing simultaneous measurements using the same local quantum sensor is a new possibility with large-spin particles that is particularly interesting, for example, because the measurements can then benefit from common noise rejection (see for example \cite{Campbell2017FermiDegClock, Savoie2018InterleavedInterf, Schioppo2016DualYbClock, Cheiney2018ContinuousAccelero}) or because it becomes possible to perform correlations analyses and thus learn from sources of noise (we will show an example in the following). Here, we present simultaneous measurements of two fields in each experimental realization. We drive two $s=1/2$ Ramsey interferometers in parallel, using two independent sets of pairs of Zeeman states within strontium's ground-state hyperfine manifold. Our scheme using sets of qubits \textit{internally distributed} within the ground-state spin manifold has similarities with the architecture of \textit{spatially distributed} interferometers studied in Ref.~\cite{Gessner2020msf}.

\begin{figure}[t!]
\centering
  \includegraphics[width=\columnwidth]{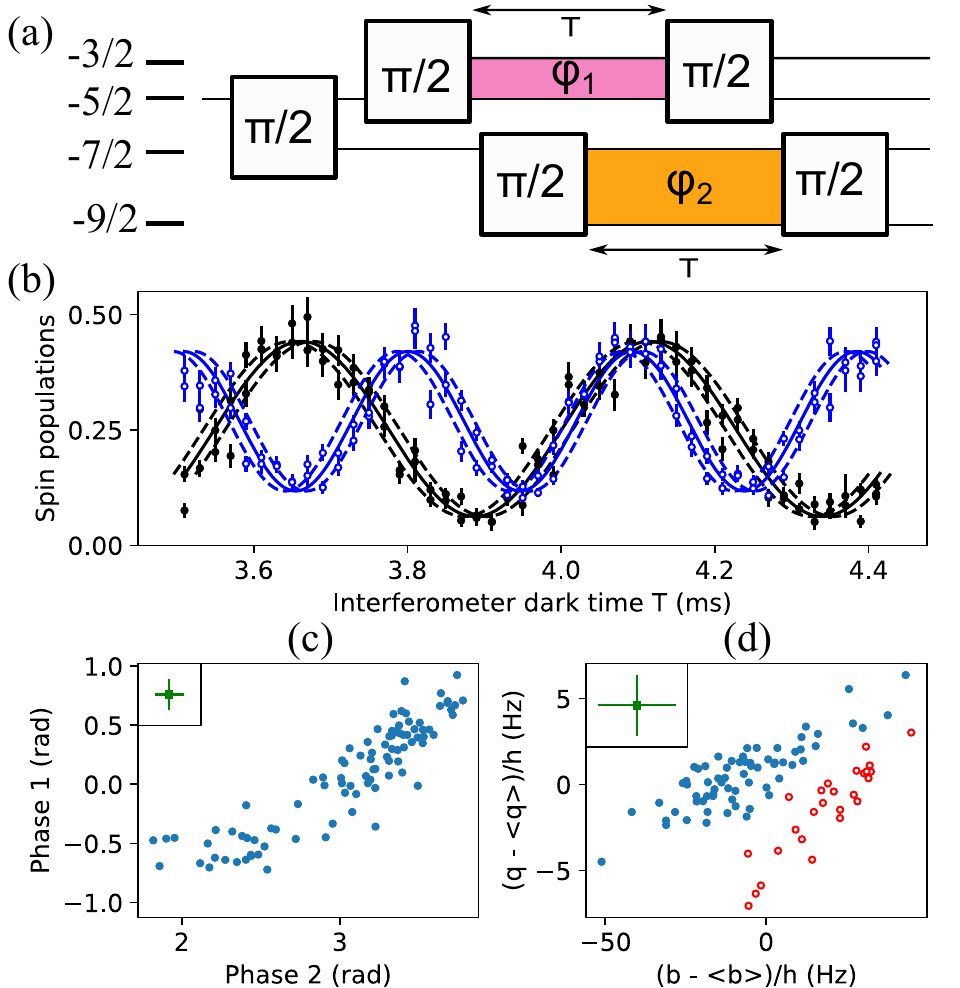} \\ 
  \caption{\textbf{Simultaneous sensing of two external fields.} a) Double interferometer pulse scheme. b) Measured interferogram. We vary the phases $\phi_1$ and $\phi_2$ together by tuning the interferometers' phase accumulation times $T$. We measure both states $\ket{-3/2}$ (black, full circles) and $\ket{-7/2}$ (blue, empty circles) in each realization.
  The differing periods highlight that two different energies are measured. The error bars combine population measurement uncertainties and the expected impact of Rabi pulse area fluctuations. The solid line is the underlying sinusoidal interferogram, barring phase fluctuations. The dashed-line interferograms represent shifts of $\pm$ one standard deviation of the phase fluctuations. 
  c) Correlation analysis of the two phases $\phi_1$ and $\phi_2$, measured modulo $2\pi$ by repeating experiments at a chosen phase with the highest slope for both interferometric signals. We show the median of the error bars (including measurement uncertainty and pulse area fluctuations) in the top-left corner. 
  d) Reconstruction of the two fields $(b,q)$. The data split into two groups, identified by color and predominantly separated in $b$ (the split corresponds to {$\phi_2 \lessgtr 2.8$\,rad}). We cancel this experimental fluctuation for most data in the paper, including Fig.~\ref{fig:5pulseInterf_clean}b, by protecting the TLS optical fiber from airflow. 
 }
 \label{fig:5pulseInterf_clean}
 \end{figure}

The protocol is shown in Fig.~\ref{fig:5pulseInterf_clean}a. From an initially pure spin state in $m_F = -5/2$ we apply a first $\pi/2$ pulse that creates a coherent superposition equally distributed over $\{-7/2, -5/2\}$ \cite{notemixture}. Next, each state enters the respective input port of two independent Ramsey interferometers, one engineered between $\{-5/2, -3/2\}$ extracting a phase $\phi_1$, the other between $\{-7/2, -9/2\}$ extracting a phase $\phi_2$. The experiment relies on a single RF local oscillator that controls the Raman detuning $\delta_{RF}$. The two interferometers are thus opened sequentially at times $t_1$ and $t_2$ by switching the Raman detuning from one resonance to the other, after which they accumulate phase for a time together, then are finally closed sequentially again. Both interferometers are open for the same duration $T$. The first interferometer's populations are thus sensitive to the phase 
\begin{eqnarray}
   \phi_1 &=& \frac{1}{\hbar} \int_{t_1}^{t_1+T} \left( E_{-5/2} - E_{-3/2} \right) dt - \int_{t_1}^{t_1+T}  \delta_{RF}(t)  dt \nonumber \\
  ~ &=& \frac{T}{\hbar} (4 q - b) - T \langle \delta_{RF} \rangle_1, 
\end{eqnarray}
and the second interferometer's populations, to 
\begin{eqnarray}
   \phi_2 &=& \frac{1}{\hbar} \int_{t_2}^{t_2+T} \left( E_{-9/2} - E_{-7/2} \right) dt  - \int_{t_2}^{t_2+T}  \delta_{RF}(t) dt \nonumber \\
   ~&=& \frac{T}{\hbar} (8 q - b) - T \langle \delta_{RF} \rangle_2.
\end{eqnarray}
Note that $\langle \delta_{RF} \rangle_1 \neq \langle \delta_{RF} \rangle_2$.
The parallel operation of the Ramsey interferometers provides a measurement of both terms contributing to the single-atom ground-state energies, that are, the linear energy splitting $b$ (due to both the magnetic Zeeman effect and the vector light shift) and the quadratic energy splitting $q$ (due to the tensor light shift). As $\langle \delta_{RF} \rangle_1$ and $ \langle \delta_{RF} \rangle_2$ are controlled (here with an accuracy below 0.02\,Hz), the difference between the two interferometers' phases $\phi_1-\phi_2$ provides a measurement of $q$, which is insensitive to shot-to-shot fluctuations in $b$, while $2 \phi_1 - \phi_2$ provides a measurement of $b$, which is insensitive to shot-to-shot fluctuations in $q$\,\footnote{Technically, $q$ and $b$ can also be time-varying during the interferometer sequence, and the interferometers are sensitive to their time-averages. Fluctuations of $(q, b)$ at the timescale of $T$ will be properly retrieved by the combined interferometers provided $T \gg 1/\Omega_R$, or if an additional laser frequency is used to engineer the Raman couplings simultaneously.}. 

In Fig.~\ref{fig:5pulseInterf_clean}b, we present the interferograms obtained by measuring the variation of the populations in $m_F = -7/2$ and $m_F = -3/2$,  relative to the total atom number, as a function of the interferometer time $T$. Here, $q / h \simeq \SI{-320}{\hertz}$ (independently deduced by spectroscopy)
and {$\Omega_R/2\pi = 77$\,Hz} on the $(-5/2 \rightarrow -7/2)$ line. Both populations are measured simultaneously at each experimental realization. During the time when both interferometers are simultaneously opened ($\simeq T - 3.5\,$ms), we set $\delta_{RF}$ to $2 \pi \times 1\,$Hz, off-resonant with both transitions. We observe that the periods of the two fringe patterns strongly differ, as $E_{-5/2} - E_{-3/2} \neq E_{-9/2} - E_{-7/2}$. From the two periods, we can deduce the mean quadratic and linear Zeeman splittings: $\langle q \rangle / h = \SI{-303(8)}{\hertz}$ and $\langle b \rangle / h = \SI{1000(45)}{\hertz}$ - in agreement with independent estimates from Raman spectra. These are mean values over all realizations, as $q,b$ can fluctuate from shot to shot. Upon knowing the fringe contrast and the two phases $\phi_1,\phi_2$ with better than $\pi$ accuracy, single realizations of the experiment provide parallel individual simultaneous measurements of $\phi_1$ and $\phi_2$, from which we can deduce single-shot measurements of $q$ and $b$. Given the interrogation time $T \simeq 4\,$ms and noise level of these data (affected by Rabi pulse area fluctuations, mostly of the first pulse), the respective precisions in $q,b$ from individual points when at maximal slopes are {$h \times 1.8$\,Hz} and {$h\times 11$\,Hz}.

To illustrate this, we operate the interferometer at a fixed $T$ chosen to be close to the maximum phase sensitivity of the interferometer ($i.e.$, close to mid-fringe for both signals). Fig.~\ref{fig:5pulseInterf_clean}c shows a set of 93 measurements of the two phases $(\phi_1, \phi_2)$ taken consecutively. For each measurement, we directly reconstruct the deviations of $b$ and $q$ relative to the mean, as shown in Fig.~\ref{fig:5pulseInterf_clean}d. These data demonstrate that the simultaneous nature of the measurement makes possible correlation analyses that would otherwise be impossible if the experiment were alternating between one interferometer and the other. 

As an example of correlation analysis, we have used this data as follows. We observe that the parallel measurements of $(q,b)$ do not seem 2D-Gaussian distributed. Instead, two classes of shots are identifiable. The two groups' separation in $\langle q \rangle$ is {$-1.4(5)$\,Hz}, i.e., a small fraction (0.4$\%$) of $\langle q \rangle$ and below the spread within a single group. In contrast, the separation in $\langle b \rangle$ of {23(3)\,Hz} is significant and higher than the spread in a single group. We attribute this to an evolution of the TLS beam polarization that shifts between two configurations; and thus induces a fluctuating vector light shift with values clustering in two groups, while there should be no vector light shift for the ideal $\pi$ polarization. By protecting the TLS optical fiber from airflow, only one group remains, and the fluctuations become consistent with measurement uncertainties (also impacted by the first pulse's area fluctuation). We acquire most of the data in this paper (including Fig.~\ref{fig:5pulseInterf_clean}b) with fiber protection.

We now discuss the fundamental limit to the sensitivity of the measurements. In the experiments presented here, the dominant uncertainties on $b$ and $q$ arise from Rabi-pulse-area fluctuations and the atom number measurement precision. Nevertheless, the atomic quantum projection noise sets the fundamental limit to the achievable precision of measurements. A key question is what the tradeoff for parallelizing two measurements is. Each one uses, on average, half of the total atom number. In addition, we present in this paper a suboptimal measurement, as we measure only one output port of each interferometer at each realization. Not knowing the exact atom number fed into each interferometer in each realization increases the quantum-projection-noise limit for phase estimation by $\sqrt{3/2}$. However, in principle, we could measure all four states in each realization \cite{Taie2010roa, Stellmer2011dam,Bataille2020}. The fluctuations in the measurement of the two phases associated with the atomic quantum-projection noise are simply $\sqrt{2}$ times stronger than when using all atoms on the same interferometer. This mild disadvantage vanishes if one considers that at least two runs of the experiment would be required to measure sequentially the two interferometers.

We finally discuss the three main systematic effects that could affect the phase measurements. 
First, the pulses of the two interferometers are interleaved temporally. Consequently, between the splitting and recombining pulses of the interferometer (1), we realize one Raman pulse addressing interferometer (2), and vice-versa. This interleaving creates AC Stark shifts $\simeq \Omega_R^2/8q$ on the two levels probed by interferometer (1) of the same sign but not exactly compensating due to different Clebsh-Gordan coefficients in the Raman couplings. These shifts can be calculated, or minimized by ensuring the highest possible energy scale separation $2q/\Omega_R$. Second, when the Rabi frequencies are only slightly smaller than the energy difference between states, one should account for effects beyond the rotating-wave approximation, such as the Bloch-Siegert shift, and expect the interferometer to be sensitive to the initial temporal phase of the beat between the two Raman beams \cite{Bidinosti2022gat}. Third, in a scheme spanning many pulses and Zeeman states, off-resonant transfers between spin states can interfere with the resonantly driven population transfers. We will see in the next section evidence for this. Effectively, this affects the fringe patterns and phase estimation.

We have shown in this section an example of a sensor that parallelizes two measurements using an extended Hilbert space basis. Thanks to the simultaneity of the measurements, we have demonstrated that correlation analyses are possible. We illustrated it by tracking in parallel variations of the vector and tensor light shifts acting on the atoms. One can also interpret these experiments as a common-mode noise-rejection scheme, i.e., extracting precise values of $q$ despite the fact that fluctuations of $b$ affect all interferometers. Similar schemes for the simultaneous sensing of multiple fields could be devised, e.g., for measuring multiple spatial components of a vector (magnetic) field.

\subsection{Simultaneous measurements of multiple collective atomic observables.}
\label{SubSec:SimultMeasurementObservables}

We will now explore a second possibility offered by the large spin-state manifold: increasing the number of observables accessible in one realization and characterizing the collective spin state of a quantum many-body ensemble. 

Let us start by remarking that tracking two spatial components of a magnetic field by monitoring spin precession requires measuring the projection of the collective spin along at least two directions simultaneously. However, this is typically not possible due to the non-commutativity of the two spin-projection observables. More fundamentally, fluctuations of a collective spin in the two directions transverse to its mean orientation are critical to the precision of metrological applications \cite{Kitagawa1993sss, Yurke1986sas} but remain assessed along one given quantization axis at a time. Furthermore, measuring the collective spin properties enables tests of a large class of entanglement witnesses \cite{Esteve2008sae, Lewenstein2000ooe, Lucke2014dme, Hosten2016mn1}. We also point out that large-spin atoms are compelling due to numerous new scenarios involving entanglement witnesses \cite{Sorensen2001eae, Vitagliano2011ssi}. For large-spin atoms, only measuring the total population in each spin state can already reveal beyond mean-field dynamics \cite{Klempt2010pao, Lepoutre2019ooe}, and measuring collective spin fluctuations in a single axis can already be used to characterize correlations \cite{Alaoui2022mcf, alaoui2024}. Here, we will show that ensembles of large-spin atoms offer an opportunity for physics with ensembles of effectively smaller pseudo-spins: measuring two orthogonal spin-projection operators - or generically, two non-commuting observables - in one given experimental realization.

The general idea is to use a larger space of states for each atom in the measurement sequence than during the experiment that came before it, so that a higher number of observables is available. These observables (e.g., populations) in the large Hilbert space commute with each other and can be simultaneously measured; but their expectation values and even higher order statistics reflect those of non-commuting observables of the state produced by the experiment.

\begin{figure}[t!]
\centering
\includegraphics[width= \columnwidth]{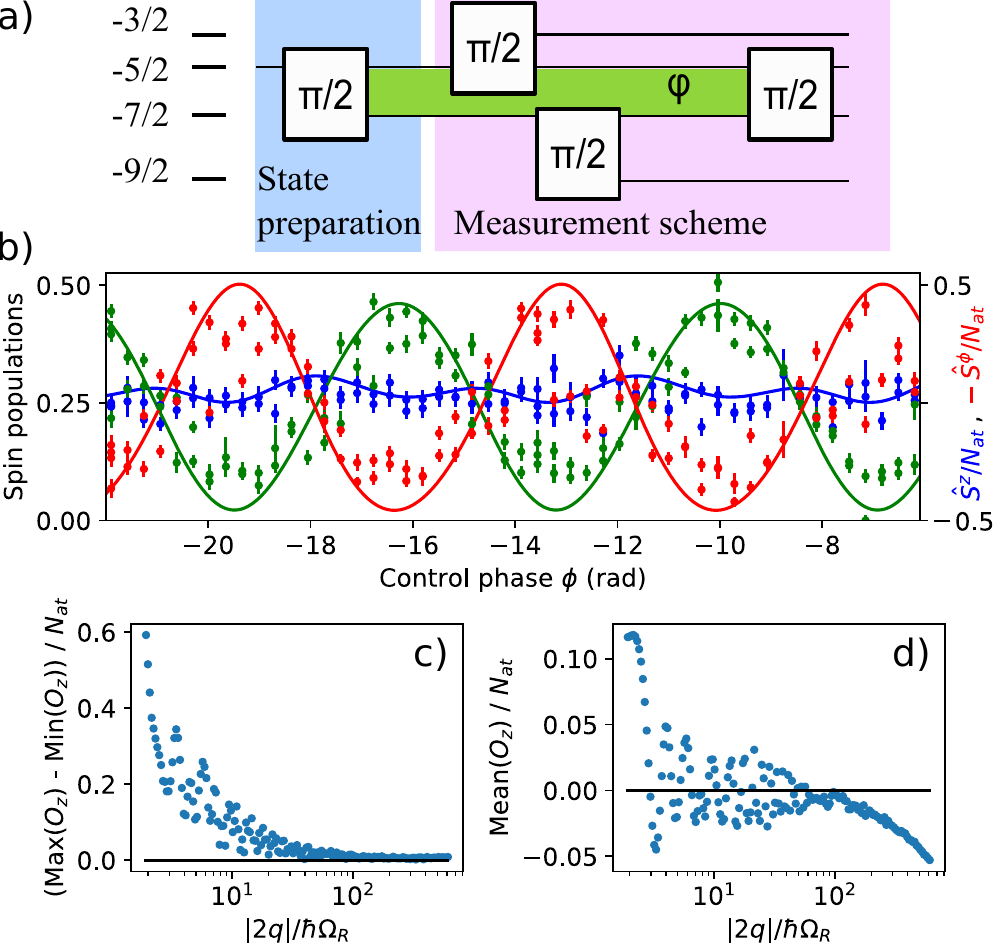}  
\caption{\textbf{Measuring two non-commuting observables in a single realization} a) Pulse scheme, highlighting one stage for preparing a well-defined state, and one measurement stage that includes unitary operations before the detection of populations. 
In frame (b), we monitor, as a function of the control phase, the fractional populations $N_{(-7/2,-5/2,-3/2)}/N_{at}$, in red, green, and blue, respectively. Error bars reflect only the population measurement uncertainty. Solid lines are the result of simulations of the master equation. We simultaneously acquire $N_{-7/2}/N_{at}$ and $N_{-3/2}/N_{at}$ during the same runs. From $N_{-3/2}$, we extract an estimate of the projection $\langle \hat{s}^z_{\uparrow, \downarrow} \rangle$ of qubit pseudo-spin, while from $N_{-7/2}$, we extract an estimate of $\langle \hat{s}^\phi_{\uparrow, \downarrow} \rangle$, all during the same runs of the experiment. We acquire the population $N_{-5/2}/N_{at}$ by independent runs. 
The solid lines result from a master equation simulation with a single adjustable parameter: an {18\,Hz} correction on the linear Zeeman splitting. (c) and (d) present the dependence of $\langle \hat{O}_z \rangle$ on the energy scale separation $2 | q | / \hbar \Omega_R$ for Rabi pulses with a simple square temporal envelope, obtained from numerical simulations. We consider a coherent state with $\langle \hat{s}^z_{\uparrow, \downarrow} \rangle = 0$. We take the maximal, minimal, and mean values of $\hat{O}_z$ over a $2 \pi$ variation of the control phase. We operate our experiment at $2 | q | / \hbar \Omega_R \simeq 9$, providing the best balance between technically-limited decoherence and scheme fidelity.}
\label{fig:4pulseInterf}
\end{figure}

One such scheme emerges if one can coherently map the internal states involved in the experiment into other, initially empty, ancillary states used for measurement purposes only. Reference~\cite{Kunkel2019} demonstrated this idea, using the complete hyperfine structure of alkali atoms to probe spinor properties in the $F=1$ manifold. We generalize this approach to the case of ground-state alkaline-earth atoms. In practice, we consider ensembles of effective qubits, i.e., pseudo-spins 1/2, initially restricted to two Zeeman sublevels. To simultaneously measure two pseudo-spin projections, we first map each state of this qubit into ancillary spin states accessible in the $F=9/2$ manifold, using a partial coherent transfer. Then, using again the capacity to perform manipulations targeted on specific pairs of states (i.e., rotations on a sub-Bloch-sphere), we perform different basis transformations on the qubit states than on the ancillary states. The final measurement of populations thus contains information formerly inaccessible in the initial populations of the qubit states.  We show a typical sequence in Fig.~\ref{fig:4pulseInterf}a. 

For our experiments, we define the two qubit states, $\ket{\uparrow} = \ket{-5/2}$ and $\ket{\downarrow} = \ket{-7/2}$. The initial state, which we produce to test the measurement protocol, is a coherent state of qubits $\ket{\psi}^{(N_{at})} = \left( (\ket{\uparrow} - i \ket{\downarrow})/\sqrt{2}\right)^{\otimes N_{at}}$, where $N_{at}$ is the atom number. We produce it from a sample polarized in $\ket{\uparrow}$, by one $\pi/2$ Raman pulse between the qubit states. With our phase convention, this coherent state is polarized against the pseudo-spin quantization axis ``y", i.e., the total pseudo-spin of the ensemble has projections $\langle \vec s_{\uparrow, \downarrow} \rangle  = (0, - N_{at}/2, 0)$.

The measurement sequence then begins. First, two successive $\pi/2$ pulses duplicate the $\ket{\uparrow}$ amplitude to the ancillary state $\ket{a} = \ket{-3/2}$ and the $\ket{\downarrow}$ amplitude to the ancillary state $\ket{b} = \ket{-9/2}$, respectively. Second, we let the phases of all four states evolve in the presence of the TLS and magnetic field. In particular, the qubit sub-Bloch sphere effectively evolves by $\exp{\left( - i \phi \sigma^z_{\uparrow, \downarrow} /2\right)}$, where the angle $\phi$ is controlled by the duration of the rotation and by the detuning between the RF clock and the qubit transition. Third, we apply a final $\pi/2$ pulse between the two qubit states, corresponding to applying $\exp{\left( - i (\pi/2) \sigma^x_{\uparrow, \downarrow} /2\right)}$. At the end of the sequence, one would ideally measure the populations $N_i$ of all four states. The population difference $N_{a}-N_{b}$ provides an estimator for the many-body operator $\hat{s}^z_{\uparrow, \downarrow}$. 
Similarly, $N_{\uparrow}-N_{\downarrow}$ provides an estimator for $\hat{s}^\phi_{\uparrow,\downarrow} = \cos(\phi) \hat{s}^y_{\uparrow, \downarrow} + \sin(\phi) \hat{s}^x_{\uparrow, \downarrow}$, the total pseudo-spin projection in one selected direction of the plane $(x,y)$. 
Both quantities are thus accessible in a single experimental run. 

Let us describe more formally the measurement, which will enable us to discuss its statistics. We label $U$ the one-body evolution operator corresponding to the series of three $\pi/2$ pulses and phase evolution within the measurement sequence, outlined in Fig.~\ref{fig:4pulseInterf}, with $\phi = 0$ for simplicity. 
We apply the same evolution to all spins in parallel. Thus, the many-body evolution operator is factorizable onto all the modes of the atomic external degrees of freedom $\lambda$: $U^{(N)} =  \Pi_{\lambda} U_{\lambda}$.
We define two projective measurement operators, 
\begin{eqnarray}
\hat{O}^z&=&U^{(N)\dagger} (N_a-N_b) U^{(N)}, ~\rm{and} \\
\hat{O}^y&=&U^{(N)\dagger} (N_\uparrow-N_\downarrow) U^{(N)}.
\end{eqnarray}
They correspond to measuring all four atomic populations ($N_a, N_b, N_\uparrow, N_\downarrow$) after the collective evolution described by $U^{(N)}$ and computing the appropriate number differences. Assuming that before the measurement sequence, the only populated states are those of the qubit $(\uparrow,\downarrow)$, one can easily show:
\begin{eqnarray}
    \left< \hat{O}^z \right> &=& \left< \hat{s}^z_{\uparrow,\downarrow} \right> , 
    \left< \hat{O}^y \right> = \left< \hat{s}^y_{\uparrow,\downarrow} \right> , \\
     \hat{O}^z \hat{O}^y &=& \hat{O}^y \hat{O}^z.
\end{eqnarray}
Note that $\hat{O}^z$ and $\hat{O}^y$ do commute with each other, which highlights the simultaneous nature of the two measurements. Thus, measuring the four populations after the outlined sequence allows us to estimate the expectation values $\langle \hat{s}^{y}_{\uparrow,\downarrow} \rangle$ and $\langle \hat{s}^{z}_{\uparrow,\downarrow} \rangle $ in a single physical realization. Furthermore, we have shown using second quantization formalism that
\begin{eqnarray}
Var(\hat{O}^{\{y,z\}}) &=& Var(\hat{s}^{\{y,z\}}_{\uparrow,\downarrow}) + N_{at}/4,
\end{eqnarray}
provided that only the qubit levels $(\uparrow,\downarrow)$ are populated before the measurement operations described by $U^{(N)}$. It is remarkable that the scheme only introduces an additive fluctuation term $N_{at}/4$, which could then be subtracted in quadrature to estimate $Var(\hat{s}^{\{y,z\}}_{\uparrow,\downarrow})$ from the measurement of $Var(\hat{O}^{\{y,z\}})$. Note that $N_{at}/4$ corresponds to the transverse spin variance of a coherent state.

Figure~\ref{fig:4pulseInterf}b shows the measured interferograms. 
Here, $q / h = \SI{-330}{\hertz}$ and $\Omega_R/2\pi = 76$\,Hz on the $(-5/2 \rightarrow -7/2)$ line. We control the phase $\phi$ by varying the Raman detuning during a fixed span of {0.51\,ms} just before the last Rabi pulse. Ideally, one would simultaneously measure the populations in all four states\,\cite{Taie2010roa, Stellmer2011dam}. 
In our setup, only two spin states, $m_F =-7/2~(\downarrow)$ and $m_F = -3/2~(a)$, can be measured in the same experimental run, as it is the easiest for our spin-population measurement protocol \cite{Bataille2020}. We measure the spin state $m_F =-5/2~(\uparrow)$ separately, whereas, in the current implementation, the Clebsch-Gordan coefficient is too weak to efficiently measure the state $m_F =-9/2$. 
Since we operate in the regime where the Rabi pulses are well resolved, we can consider that $N_{-9/2}+N_{-3/2} = N_{at}/2$ and $N_{-7/2}+N_{-5/2} = N_{at}/2$ and thus relate each population to a spin observable,
\begin{eqnarray}
 \langle  \hat{s}^z_{\uparrow,\downarrow} \rangle & =& \langle U^{(N)\dagger} \left(2 N_{-3/2} - N_{at}/2 \right) U^{(N)} \rangle, \\
  \langle  \hat{s}^\phi_{\uparrow,\downarrow} \rangle & =& - \langle U^{(N)\dagger} \left( 2 N_{-7/2} - N_{at}/2 \right) U^{(N)}\rangle.
\end{eqnarray}
However, the fluctuation of these observables will be higher than those of $\hat{O}^z$ and $\hat{O}^\phi$, which are more optimal if all four states can be measured simultaneously. 

The striking feature of Fig.~\ref{fig:4pulseInterf}b is that the populations in states $m_F= -7/2~(\downarrow)$ and $-5/2~(\uparrow)$ show pronounced oscillations, while the population in state $m_F =-3/2~(a)$ barely evolves. The oscillations reflect that we effectively change the spin measurement basis with the interferometric phase. The approximately flat behavior of $N_{-3/2}$ demonstrates the capacity to rotate the spin in the sub-Bloch sphere set by the states $\ket{-7/2}$ and $\ket{-5/2}$ without significantly modifying the population $N_{-3/2}$. We still observe a weak dependence of $N_{-3/2}$ on the control phase $\phi$. This signal is a signature of population transfers driven off-resonantly by the Raman couplings, which will be finite as long as $\left| \hbar \Omega_R / 2 q \right| >0$. Interestingly, we did not see a signature of off-resonant transfers in the experiments restricted to a single pair of states. Indeed, in more complex pulse schemes, ``leaked" atomic amplitudes have the opportunity to interfere with macroscopic amplitudes that are resonantly driven, which exacerbates their impact on populations. 

The simple approach to tackle these parasitic population transfers is to increase the energy scale separation, $\left|  2 q / \hbar \Omega_R  \right|$. We show in Fig.~\ref{fig:4pulseInterf}c,d how the stray signals fall with reduced Rabi couplings, using numerical simulations of the master equation. Here, the fundamental source of dissipation ($i.e.$, spontaneous emission due to a perfectly monochromatic laser) is included. The studied quantity is $\langle \hat{O}_z \rangle / N_{at}$ for a coherent state with $\langle \hat{s}^z_{\uparrow, \downarrow} \rangle = 0$. Provided the phase control is achieved in a time shorter than the Raman pulses, this quantity depends on the ratio $ 2 |q| / \hbar \Omega_R$, irrespective of the values of $q$ and $\Omega_R$. We see that the $\phi$-dependence and mean value of $\langle \hat{O}_z \rangle / N_{at}$ both slowly approach the quantum-projection-noise limit ($\sqrt{ 1 /2 N_{at}}  = 0.007$ for $10^4$ atoms) when increasing $ 2 |q| / \hbar \Omega_R$. However, at values $ 2 |q| / \hbar \Omega_R > 10^2 $, spontaneous emission introduces spin diffusion and degrades the scheme. Technical imperfections will bring the optimum to faster pulses, i.e., lower values of $ 2 |q| / \hbar \Omega_R$.

To push this limitation, a most promising avenue is the temporal shaping of the Rabi pulses. By deviating from the square amplitude envelopes used here, even with simple shapes\,\cite{Leprince2024}, one can drastically reduce off-resonant population transfers for a given Raman coupling strength, and thus at constant overall scheme duration. More dramatically, Ref.\,\cite{Omanakuttan2021} demonstrates numerically that optimal control methods would enable performing arbitrary unitary $\mathfrak{su}$(10) operations in a time $\sim 30 \hbar / |q|$. This is about the same duration as the experiments that we present here, and relies on high inter-state couplings: $2 |q| / \hbar \Omega_R \lesssim 1$. One could also use the ultra-narrow transition \ThreePTwoTransition~\,\cite{Onishchenko2019} to engineer the tensor light shift with dramatically reduced spontaneous emission.

Thus, expanding the internal Hilbert space in the measurement sequence enables the parallel estimates of several atomic observables. This follows the idea exposed in \cite{Peres1991} for implementing positive-operator-valued measures (POVMs), and the literature on POVMs, e.g. to estimate measurement fluctuations \cite{Massar2007}, applies. Here, we have studied the case where two orthogonal pseudo-spin projections of an ensemble of qubits are measured by expanding the space to four states. This expansion comes at the cost of an increased impact from the quantum-projection noise, owing to smaller populations in each measured state and the stochastic nature of the ``mapping" pulses. Given that there are ten possible spin states in strontium's ground state, one could, in principle, simultaneously measure all three spin projections of an ensemble of pseudo-spins $s=1/2$ or even $s=1$. Alternatively, one could simultaneously measure two projections of $s=1/2$, $s=1$, $s=3/2$, or $s=2$ pseudo-spin ensembles.

 \section{Conclusion}
 \label{Sec:Conclusion}

In this work, we have demonstrated the implementation of operations effectively driven by generators of the $\mathfrak{su}$(10) group. We focused on selectively driving Rabi oscillations between adjacent ($m_F \longleftrightarrow m_F+1$) and next-to-adjacent ($m_F \longleftrightarrow m_F+2$) levels. We demonstrated these on four out of ten levels, with high fidelity $>0.99$. We expect the generalization to the entire spin manifold to require moderate technical adjustments, such as a higher magnetic field bias. Being able to drive all $\Delta m_F = (1,2)$ transitions effectively amounts to driving ``rotations" around $2N-3 = 17$ generators of the $\mathfrak{su}$(N) group. This number exceeds the $N-1$ number of generators that can be simultaneously diagonalized (which  correspond to the $N-1$ quantization axes). Therefore concatenating $\Delta m_F = (1,2)$ transitions is in principle sufficient to implement all the $N^2-1$ generators. 

In practice, this set of rotations is sufficient to measure the relative amplitudes and coherences between all pairs of Zeeman levels, i.e perform tomography and thus reconstruct the complete state. Combining these schemes with a spin-selective detection capable of giving access to all ten spin components, as in Refs.~\cite{Stellmer2011dam, Taie2010roa}, one would gain comprehensive control of the spin state, with applications to quantum simulation, computation, or sensing schemes requiring operations within the $\mathfrak{su}$(N) group \cite{Kiktenko2015sqr, Omanakuttan2021, Godfrin2018gri, Randall2015epa, Lanyon2008sql}. Thanks to these tools, the spins of strontium 87 atoms are a fully exploitable qudit resource with $\mathfrak{su}$(10) character, which one can even complement by other coherently controllable degrees of freedom in electronically-excited states, e.g., as in Ref.~\cite{Pucher2024, Nakamura2024}.

These assertions also apply to species with similar atomic structures - alkaline-earth-like atoms and group IIB transition metals fermionic isotopes - to which the scheme could be implemented. Indeed, they have in common: a large ratio between the hyperfine splitting and the linewidth of the state $^3P_1$, favorable for tensor-shift engineering with low spontaneous-emission levels; a vast spin Hilbert space in the $^1S_0$ ground state; $\mathfrak{su}$(N) symmetric interactions; extensive coherence times afforded by the Land\'e factor of nuclear spins.

We have seen that large-spin atoms open up the possibility of parallelized interferometers, which can be used for measuring multiple fields or for common-mode noise-rejection schemes. Similar sequences could readily explore other avenues, such as multiple-path interferometers and quantum random walks. More generally, being able to tailor the spin state by tensor light shifts enables the production of ``non-classical" spin states, i.e., single-atom spin states that cannot be modeled by an assembly of uncorrelated spin-1/2 particles \cite{Satoor2021}. These ``non-classical" spin states present specific advantages for quantum sensing\,\cite{Chalopin2018qes}. 
 
Finally, following the original idea of Ref.~\cite{Kunkel2019}, we have seen that the vast Hilbert space is an opportunity for the parallel estimation of multiple non-commuting observables of an atomic ensemble. We studied this in the case of two observables of an ensemble of pseudo-spins 1/2. However, it would be most interesting to study the parallel estimation, e.g., of three non-commuting observables or pseudo-spins $>1/2$. It would come at the cost of more Hilbert space resources and increased quantum-projection-noise effects. 
The full potential of such measurements deserve investigation. The scheme is analogous to randomly partitioning the system into several sub-ensembles and measuring each along a different quantization axis. 
One question is whether this could be amenable to extracting signatures of specific many-body states (chiral states, cluster states \cite{Bornet2024, Mamaev2019}), or to characterize general properties of the many-body states (entanglement, purity) in the spirit of, e.g., Refs.~\cite{Brydges2019, Frerot2021}.

\begin{acknowledgments}
\medskip \noindent
We acknowledge Pierre Bataille for early contributions to this work, and Alban Meyroneinc for participation in the $\delta m_F = 2$ Rabi oscillation experiments. This research was funded by Agence Nationale de la Recherche (project ANR-23-CE47-0006 LOQUST), the Conseil R\'egional d'Ile-de-France, DIM Sirteq (projects SureSpin and Suprisa), DIM Quantip (projet structurant 2022), and by the PHC Polonium program (2024 project 50998RL). Pauline Guesdon acknowledges funding by the Program QuanTEdu-France ANR-22-CMAS-0001 France 2030.

\noindent
\textbf{Correspondence and requests for materials} should be addressed to M.R.D.S.V. 

\end{acknowledgments}


\begin{thebibliography}{76}
\bibitem [{not()}]{notemixture}%
  \BibitemOpen
  \href@noop {} {}\bibinfo {note} {We could equally have used an incoherent
  mixture of these two states.}\BibitemShut {Stop}%
\end{thebibliography}

%

\end{document}